# *Tailoring porous media for controllable capillary flow*

Mingchao Liu[a, b], Si Suo[b], Jian Wu[a], Yixiang Gan[b], Dorian AH Hanaor[c] and C.Q. Chen[a, *]

a. Department of Engineering Mechanics, CNMM & AML, Tsinghua University, Beijing 100084, China
b. School of Civil Engineering, The University of Sydney, Sydney, NSW 2006, Australia
c. Chair of Advanced Ceramic Materials, Technische Universität Berlin, Berlin 10623, Germany
*. Corresponding author. Tel/Fax: +86 10 62783488; Email: chencq@tsinghua.edu.cn

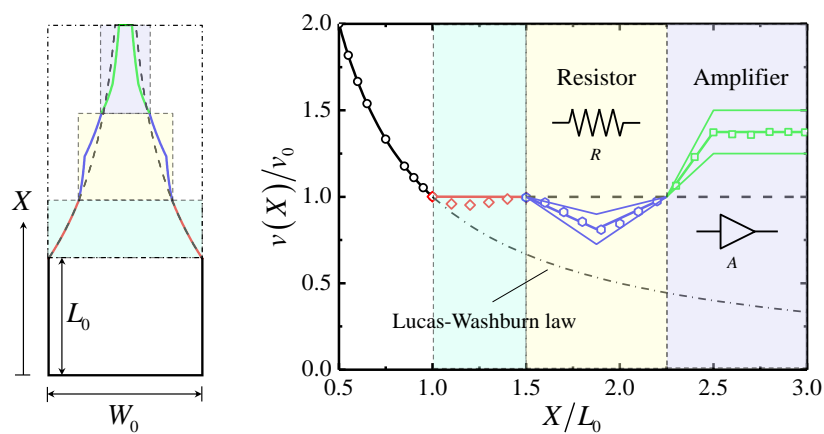

Abstract

*Hypothesis*
Control of capillary flow through porous media has broad practical implications. However, achieving accurate and reliable control of such processes by tuning the pore size or by modification of interface wettability remains challenging. Here we propose that the flow of liquid by capillary penetration can be accurately adjusted by tuning the geometry of porous media and develop numerical method to achieve this.

*Methodologies*
On the basis of Darcy's law, a general framework is proposed to facilitate the control of capillary flow in porous systems by tailoring the geometric shape of porous structures. A numerical simulation approach based on finite element method is also employed to validate the theoretical prediction.

*Findings*
A basic capillary component with a tunable velocity gradient is designed according to the proposed framework. By using the basic component, two functional capillary elements, namely, (i) flow amplifier and (ii) flow resistor, are demonstrated. Then, multi-functional fluidic devices with controllable capillary flow are realized by integrating the designed capillary elements. All the theoretical designs are validated by numerical simulations. Finally, it is shown that the proposed model can be extended to three-dimensional designs of porous media.

## Highlights

1. A theoretical method is developed to design porous structures with controllable flow.
2. Two novel capillary flow elements, i.e., amplifier and resistor, are designed.
3. Controllable capillary flow is achieved by integrating the designed capillary elements.
4. The extension of the developed method to 3D conditions is discussed.

## 1. Introduction

Capillary-driven flows in porous media are governed by multi-scale interfacial interactions of gases, liquids and solids and consequently represent a challenging aspect of engineering problems. Fluid flow in porous media, also referred to as imbibition, is frequently encountered in diverse contexts ranging from natural systems, such as flow in biological tissues [1] and rising damp in architectural structures [2], to industrial settings including paper-based microfluidics [3], medical diagnosis [4], energy-harvesting devices [5], and oil recovery [6]. Following a rapid expansion in the application of paper-based microfluidic devices, the ability to control and promote capillary fluid flow in porous media is of increasing interest in recent years [7], particularly as such control can facilitate the improvement of chemical sensing and medical diagnostic devices [8, 9].

The study of liquid flow in capillary tubes has a rich history. Pioneering work dates back to Lucas [10] and Washburn [11], who over a century ago established a diffusive relationship between the flow distance $L$ and time $t$ (i.e., $L^2=D \cdot t$, where $D$ is the diffusive coefficient depending on the tube size, liquid properties, and the interface properties between liquid and the tube wall [12]). This law is valid when the effect of gravity can be ignored [13], and has been extended to characterize capillary flow in converging [14] and tortuous tubes [15]. In comparison with the capillary rise in a hollow tube, the capillary flow in porous media is much more complex, but shares a similar dynamic mechanism, related to the effective pore radius [16].

As similar dynamics are involved, the Lucas-Washburn law has been applied to describe the unidirectional capillary flow in homogeneous porous media [16], and has also been modified to analyze non-unidirectional (i.e., radial and hemispherical) capillary flow [17, 18]. Moreover, several fundamental and practical aspects of capillary flow in porous media have been the subjects of recent research efforts, including the effects of liquid evaporation [19, 20], gravity [21], swelling of the porous matrix [22, 23], and fractal and disordered microstructures [24, 25]. Additionally, the capillary flow has been utilized as an inverse method to infer the effective properties of porous media, such as the pore size distribution and porosity [26].

In recent years, several approaches have been employed to control fluid flow in porous media, including the compression of paper to decrease its average pore size [27], surface treatments to alter wettability [28], and the use of packed beads to achieve graded permeability [29]. However, accurate, consistent and robust modification of surface interactions in porous materials is problematic, particularly in biomedical applications, and most porous media are not conducive to the control of fluid flow by variation of pore size or porosity [30]. In contrast, geometry-based methods for flow velocity control present a simpler and more robust approach [7].

For the purpose of facilitating the control of fluid flow, porous media with variable cross-sections in the direction of flow are commonly used [7]. In earlier studies, simple structures with variable geometries were considered, including triangular shaped blotting paper and multi-sectional porous media [31, 32]. It has been shown that the time dependence of liquid front motion in expanding porous media deviates from the one-dimensional Lucas-Washburn law [33] and by means of a numerical method, flow in complex-shaped and heterogeneous porous media were investigated [34, 35]. Recently, the effects of gravity and evaporation, have been incorporated into approaches to flow-control design [36, 37]. In addition, it has been shown that the velocity of capillary flow in capillary tubes and porous media can be regulated by employing concepts of capillary elements, in the form of flow diodes, resistors, and keys [38-41].

Most aforementioned studies are on the capillary flow in porous systems with fixed geometries, with a few focusing on the geometric design to produce a targeted fluid velocity distribution. Regulating capillary-driven flows in paper-based microfluidics has been the subject of recent studies in which the shape of a porous structure yielding constant liquid front velocity was designed and validated by experiments [42]. An iterative computational optimization approach has also been proposed for the design of porous structures with constant flow velocity and constant volumetric flow [43]. However, the inverse design of porous structures with the generally controllable capillary flow, to the best of our knowledge, has yet to be thoroughly explored and is the objective of the present work.

Here we introduce a framework to achieve a controllable flow velocity by tailoring the cross-sectional profile of homogeneous porous media. Within this framework, a basic capillary flow component is designed to achieve a flow process with constant velocity gradient in the sample. On this basis, two typical functional capillary elements, namely flow amplifiers and resistors, are designed by tailoring the cross-sectional shape of multi-sectional porous media. Moreover, a simple example of a fluidic device with controllable capillary flow process in porous media is achieved by integrating the designed capillary elements. Additionally, in order to verify the theoretical design of the geometrical profile of shaped porous media for the purpose of controlling capillary flow, a numerical simulation based on a finite element method is developed to simulate flow in shaped porous media. Even though the design of the porous structures presented involves two-dimensional (2D) conditions, the adaptation of this methodology for three-dimensional (3D) structures is also discussed.

## 2. Basic concept

A generalized homogeneous porous medium consisting of multiple sections with uniform thickness $T$ is considered in this study (see Fig. 1(a)). To facilitate the analysis, two sets of coordinates are considered. The global coordinate, $X$, starting from the end of the rectangular section, and a series of local coordinates, $x_i$, indicating the $i^{th}$ variable section from input to output ends, where $i = 1 \ldots n$. As shown in Fig. 1(a), the length of each section is $L_i$. The initial rectangular section with length $L_0$ and uniform width $W_0$ serves to supply the hydrodynamic load [42, 43], followed by multiple sections with variable widths along the $x_i$ direction as $w_i(x_i)$, in order to provide the targeted flow. The expression of $w_i(x_i)$ depends on the prescribed flow velocity along the flow direction and the initial value at each section denoted by $w_i(0) = W_{i-1}$ according to the condition for continuity. Owing to its uniform thickness, flow in this porous medium is uniform in the transverse direction.

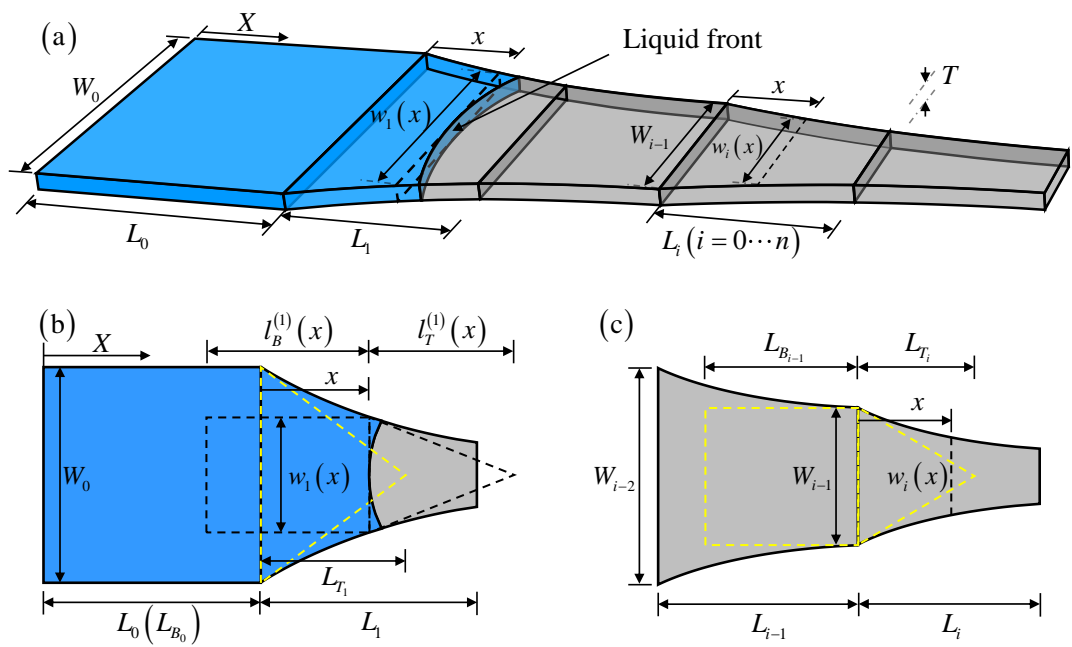

**Figure 1.** Schematics of a porous medium with variable cross sections: (a) Oblique view of the whole structure with multiple sections; (b) Top view of the first two sections; (c) Top view of any two adjacent sections.

It should be noted that the liquid front in non-uniform porous structures is generally not a straight line perpendicular to the flow direction, as considered in a one-dimensional (1D) model, but rather an arc, as shown in Fig. 1(a). Therefore, this case should be regarded as a 2D problem [33]. However, previous studies have indicated that the relative error of the 1D model with the assumption of a flat liquid front, in comparison with the 2D model, is comparatively small for moderate opening angles [37, 42]. Therefore, to simplify the analysis, we adopt the 1D assumption of a flat liquid front (see the dashed line in Fig. 1(a)) to establish the theoretical model through which geometric control of fluid flow is imparted.

## 3. Theoretical background

To design the geometric shape of a porous structure with a desired distribution of velocity in the flow direction, the relationship between the segment width and the flow velocity should be obtained. Here, we first consider a porous structure combined with a rectangular supply sample and a variable section, see Fig. 1(b). The lengths of rectangular and variable sections are $L_0$ and $L_1$, respectively. The width of the rectangular section is $W_0$, which is also the width of the input end of the variable section, i.e., $w_1(0) = W_0$. At the edge of the variable section, a triangle is constructed with hypotenuse tangents to the curved boundary (see the yellow dash-dot line in Fig. 1(b)). The length of the midline of this triangle is $L_{T_1}$, and the width of the two-stage structure consisting of a rectangle and a triangle can be expressed as

$$w(X) = \begin{cases} W_0 & , \ 0 \leq X \leq L_0 \\ W_0 \cdot \left(1 + \dfrac{L_0 - X}{L_{T_1}}\right), & L_0 \leq X \leq L_0 + L_{T_1} \end{cases}, \quad (1)$$

where $X$ is the global coordinate along the flow direction, with the original point at the end of the rectangular section, as shown in Fig. 1(b).

Capillary flow is governed by properties of the gas, liquid and solid phases in the media. Considering Darcy's law and the conservation of mass, the governing equation of the liquid front at the variable section can be expressed as [37]

$$P_c = \frac{\phi \mu}{k}(L_0 + L_{T_1} - l)\left[\frac{L_0}{L_{T_1}} + \ln\left(\frac{L_{T_1}}{L_0 + L_{T_1} - l}\right)\right] \cdot \dot{l}, \quad (2)$$

where $\mu$ is the viscosity of the liquid, $\phi$ and $k$ are the porosity and permeability of the porous medium, respectively, and $P_c = 2\sigma \cos\theta_S / R_{eff}$ is the capillary force, which is also the pressure difference between reservoir ($P_{atm}$) and the liquid front ($P_{atm} - P_c$), depending upon the air–liquid surface tension $\sigma$, the equilibrium contact angle of the liquid with the solid $\theta_S$, and the effective pore radius of the porous medium $R_{eff}$. The expression $\dot{l} = dl/dt$ corresponds to the liquid front velocity, with $l$ being the liquid front position in the global coordinate system (i.e., the length of the wetted region measured from the beginning of the reservoir).

It should be noted that Eq. (2) is valid in the range of $L_0 \leq l \leq L_0 + L_{T_1}$, and the velocity $\dot{l}$ varies with position for a given geometric shape, and the velocity field is not uniform along the flow direction at a specific time. However, in the present work, we focus only on the velocity at the liquid front.

According to Eq. (2), the liquid front velocity at the input end of the variable section, i.e., $l = L_0$, can be obtained as

$$v_0 = \dot{l}\big|_{l \to L_0} = D/L_0 \quad , \tag{3}$$

where $D = P_c k / \phi \mu$ is the diffusion coefficient. Moreover, the gradient of the liquid front velocity with respect to position (i.e., the velocity gradient), at the input edge of the variable section, can be obtained by reordering Eq. (2) and deriving $\dot{l}$ to $l$ as

$$\frac{\partial \dot{l}}{\partial l}\bigg|_{l \to L_0} = D \cdot \frac{L_0 - L_{T_1}}{L_0^2 L_{T_1}} \quad . \tag{4}$$

By setting $L_{T_1} = \alpha_1 \cdot L_0$ in which $\alpha_1$ is a non-dimensional factor, Eq. (4) can be re-arranged as

$$a_1 = D \cdot \frac{1 - \alpha_1}{\alpha_1} \cdot \frac{1}{L_0^2} \quad , \tag{5}$$

where $a_1 = \partial \dot{l}/\partial l$ is the gradient of the liquid front velocity at the input end of the variable section. This parameter describes the velocity distribution along the liquid flow direction. It is clear that the sign of the velocity gradient $a_1$ depends on the value of $\alpha_1$, i.e., $0 < \alpha_1 < 1$ denotes $a_1 > 0$, corresponding to a positive gradient (here we refer it as acceleration, more precisely, velocity increases with position); while $\alpha_1 < 0$ or $\alpha_1 > 1$ denotes $a_1 < 0$, corresponding to a negative gradient (deceleration, i.e., velocity gradually decreases); $\alpha_1 = 1$ is found when $a_1 = 0$, and a uniform velocity is present.

Assuming the liquid front velocity gradient $a_1$ is constant for $x_1$ in the variable section, the corresponding liquid front velocity can be written as

$$v_1(x_1) = v_0 + a_1 \cdot x_1 \quad . \tag{6}$$

Substituting Eqs. (3) and (5) into Eq. (6), the velocity $v_1(x_1)$ can be related to the length of the rectangular section $L_0$ by the following expression

$$v_1(x_1) = D \cdot \left( \frac{1}{L_0} + \frac{1-\alpha_1}{\alpha_1} \cdot \frac{x_1}{L_0^2} \right) . \tag{7}$$

At any position $x_1$ of the variable section, an isosceles triangle can be found with its two equal sides tangent to the boundary curves, as shown in Fig. 1(b) by a dotted line, with a median length of $l_T^{(1)}(x_1)$. At the input end of the first variable section, i.e., $x_1 = 0$, a contracted notation is given as $l_T^{(1)}(0) = L_{T_1}$. Subsequently, a rectangle with length $l_B^{(1)}(x_1)$ is connected to the triangle vertex, with $l_B^{(1)}(0) = L_{B_0}$, meaning $L_{B_0} = L_0$.

Analogous to the analysis of $v_0$ from the combined structure with rectangular and triangular sections (i.e., Eq. (3)), the velocity $v_1(x_1)$ can be obtained as

$$v_1(x_1) = D/l_B^{(1)}(x_1) . \tag{8}$$

Combining Eqs. (7) and (8), we can get

$$l_B^{(1)}(x_1) = L_{B_0}^2 \left/ \left( L_{B_0} + \frac{1-\alpha_1}{\alpha_1} x_1 \right) \right. . \tag{9}$$

Moreover, as Eq. (4) is valid for any position $x_1$, we can obtain

$$\frac{a_1}{D} = \frac{l_B^{(1)}(x_1) - l_T^{(1)}(x_1)}{l_T^{(1)}(x_1) \left[ l_B^{(1)}(x_1) \right]^2} . \tag{10}$$

Substituting Eq. (9) into Eq. (10), we can get

$$l_T^{(1)}(x_1) = L_{B_0}^2 \left( \frac{\alpha_1}{1-\alpha_1} L_{B_0} + x_1 \right) \left/ \left[ \frac{\alpha_1}{1-\alpha_1} \left( L_{B_0} + \frac{1-\alpha_1}{\alpha_1} x_1 \right)^2 + \frac{1-\alpha_1}{\alpha_1} L_{B_0}^2 \right] \right. . \tag{11}$$

The geometric relationship at position $x_1$ can be expressed as

$$dx_1 = -\frac{1}{2} dw_1(x_1) \tan\theta(x_1) , \tag{12}$$

where $\tan\theta(x_1) = 2l_T^{(1)}(x_1)/w_1(x_1)$ is the tangent of the base angle of the triangle. Integrating Eq. (12), we obtain

$$\frac{w_1(x_1)}{W_0} = \mathrm{Exp}\left[-\left(\frac{x_1}{L_{B_0}} + \frac{1}{2} \cdot \frac{1-\alpha_1}{\alpha_1} \cdot \frac{x_1^2}{L_{B_0}^2}\right)\right] \bigg/ \left(1 + \frac{1-\alpha_1}{\alpha_1} \cdot \frac{x_1}{L_{B_0}}\right) . \tag{13}$$

This equation indicates the cross-sectional profile of the variable section in the porous media in terms of the capillary flow with constant liquid front velocity gradient. In other words, if we know the distribution of the liquid front velocity gradient during the flow process in advance, the geometric shape profile of the porous sample can be inversely determined. Specifically, when $\alpha_1 = 1$, Eq. (13) will reduce to a simple exponential function, which corresponds to the special case of uniform velocity, i.e., $a_1 = 0$, and is consistent with previous results [42].

According to this general model (13), we can tailor the geometric shape of the porous sample by combining multiple sections to yield a prescribed liquid front velocity. For the $i^{th}$ variable section, as shown in Fig. 1(c), one can find a triangle with midline length $L_{T_i}$, in which the hypotenuse is tangential to the curved boundary, and a rectangle with length $L_{B_{i-1}}$ is connected to the end of the triangle, see the yellow dashed and dotted lines. In the same way, based on the derivation of the above model for the first two sections, Eq. (13) can be extended to any variable section (i.e., $i = 1\ldots n$) by replacing the subscript index $1 \to i$ and $0 \to i-1$ (see Fig. 1(b)).

For each variable section, as shown in Fig. 1(c), three parameters are needed to determine the shape function $w_i(x_i)$, i.e., the initial width, $W_{i-1}$, the rectangle length, $L_{B_{i-1}}$, and the non-dimensional factor, $\alpha_i$. $L_{B_{i-1}}$ can be obtained through Eq. (9) by incorporating the geometric information of the previous section as

$$L_{B_i} = l_B^{(i)}(L_i) = L_{B_{i-1}}^2 \bigg/ \left(L_{B_{i-1}} + \frac{1-\alpha_i}{\alpha_i} L_i\right) , \tag{14}$$

where $i = 1\ldots n$. Specifically, for $i = 0$ $L_{B_0} = L_0$. The non-dimensional factor $\alpha_i$ can be determined from the given distribution of the velocity gradient at each section, $a_i$, through the following correlation

$$\alpha_i = 1 \bigg/ \left(D \cdot \frac{L_{i-1}^2}{a_i} + 1\right) . \tag{15}$$

Additionally, considering structural continuity, the initial width of each variable section is equal to that of the end of the preceding section, i.e.,

$$W_i = w_i(L_i), \tag{16}$$

where $i = 1 \ldots n$. Specifically, the width of the input end of the first variable section is equal to the width of the rectangular section, $W_0$.

For each section, substituting Eqs. (14)-(16) into Eq. (13), the cross-sectional profile of the porous structure can be precisely obtained. The procedure constitutes an explicit method to achieve a controllable flow process by inversely designing the geometric shape of homogeneous porous media. In the following sections, we will employ this methodology to implement the design of capillary elements based on porous media, and further demonstrate the realizability of controllable capillary flow by integrating the designed capillary elements.

## 4. Numerical method

As mentioned in the preceding section, a flat liquid front is assumed in the theoretical model, which is only accurate and valid for a 1D situation. In order to verify the feasibility of the theoretical design of the capillary elements and the porous samples with controllable flow features, a numerical method based on the Scaled Boundary Finite Element Method (SBFEM) [44] is employed to simulate a full 2D problem.

The 2D capillary penetration process in porous media can be controlled by a set of simultaneous partial differential equations [37], viz., Darcy's law

$$\mathbf{v} = -\frac{k_i}{\mu} \nabla P \tag{17}$$

and the conservation of mass

$$\nabla \cdot \left[ \rho \left( -\frac{k_i}{\mu} \nabla P \right) \right] = F \tag{18}$$

where $\rho$ is the fluid density, $k_i = k/\phi$ is the interstitial permeability, and $F$ is a source term. In the case of capped devices, $F$ was set to zero. Otherwise, the liquid evaporation is not negligible, and it can be calculated as $F = \dot{m}_e / H \phi$. The above Eqs. (17) and (18) govern the 2D penetration process in the full region. By solving the set of equations in a given region, one can obtain the full liquid flow velocity field, as well as the liquid front boundary of the penetration region at any time.

In the finite element analysis, the boundary conditions of the numerical model are defined such that the left and right surfaces are non-penetrable and symmetrical, and the reservoir-contacting boundary is stationary. The arbitrarily curved boundary is discretized as a series of linear segments with different slopes. Based on the continuity of the velocity field, i.e., the absolute value of velocity cannot change instantly, an appropriate flux is added to a boundary to modulate the direction of velocity along the slope of the next segment. Consequently, the whole region is solved to re-determine the velocity of the liquid front for the next time step. Moreover, a mesh sensitivity study has also been conducted a priori to ensure the convergence of the numerical models.

## 5. Design and validation of basic capillary component

As we have shown in Section 3, the flow velocity gradient can be related to the geometric shape of the tailored porous structure. For an arbitrary flow process with complex velocity distribution, it can be approximated by a number of piecewise parts with a constant velocity gradient. Here, a capillary flow sample with a constant velocity gradient is introduced. It can be used as a basic component for the design of functional capillary elements.

Consider a homogeneous porous medium with the variable cross-sectional profiles to realize the control of the flow velocity. It can be implemented by a porous medium consisting of three sections, i.e., a rectangular section appended to a variable section which is followed by another rectangular section. The middle variable section can be treated as a control section to maintain a constant gradient of flow velocity. As schematically shown in Fig. 2(a), three cases with different velocity gradients at the control section, i.e., $a = D/2L_0^2$, 0, and $-D/2L_0^2$, are considered, shown in different colors. The corresponding width distribution of the variable control section can be determined by employing the theoretical methodology, of Eq. (13) in conjunction with Eqs. (14)-(16), developed in Section 3. Specifically, the green profile represents a regulator yielding uniform velocity, and the red and blue profiles represent those imparting acceleration and deceleration, respectively.

For the purpose of validating the theoretical design of the basic capillary component shown in Fig. 2a, the flow process in a shaped porous medium under 2D conditions is simulated by using the numerical method described in Section 4. The flow processes in the components, as shown in Fig. 2(a), are simulated, and the simulated results of the flow region at a specific time (here we choose the middle point of the liquid front located at the middle position of the control section as a reference) with pressure profiles of the full field are shown in Fig. 2(c)-(e). It can be seen that the simulated liquid fronts are not straight lines for all the three cases. To simplify the comparison, we choose the average value of the position along the concave curve to represent the liquid front position and consider the average velocity to represent the liquid front velocity during the flow process.

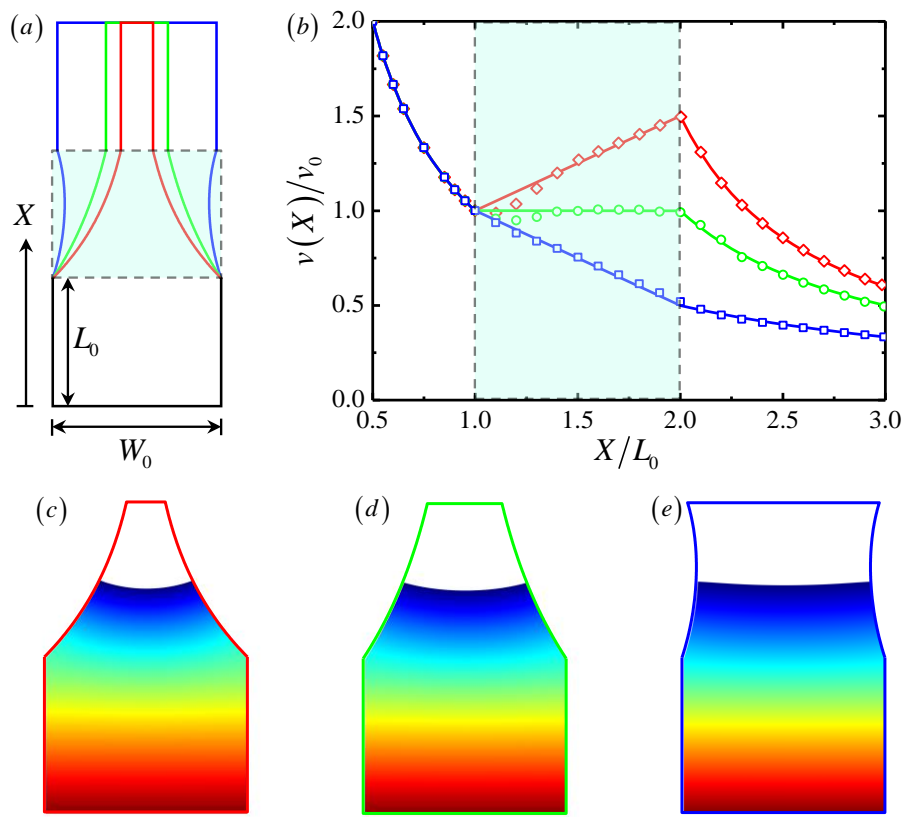

**Figure 2.** (a) Designed shape profiles of three capillary elements with different velocity gradients, i.e., red, blue and pink colors correspond to the acceleration, uniform velocity and deceleration cases, respectively; (b) Comparison of numerical simulations (symbols) and theoretical predictions (lines) of liquid front velocity versus position along flow direction; (c)-(e) The simulated results of flow region at specific time with pressure profiles for three elements, in which the contours from red to blue indicate the pressure level varying from $P_{atm}$ to $P_{atm}-P_c$.

The simulated results of the distribution of liquid front velocity in the designed components are shown in Fig. 2(b) as symbols. In addition, the position-dependent velocity predicted by the theoretical model is also included and represented by solid lines with different colors. The results are nondimensionalized by two scaling parameters $l_0$ and $v_0$, with $l_0$ being the length of the rectangular supply sample and $v_0$ the liquid front velocity at the end of the supply sample. Good agreement between the numerical and theoretical predictions on the velocity variation is obtained for all three cases (Fig. 2b). Specifically, the relative error for all the three cases (i.e., $a = D/2L_0^2$, $0$, and $-D/2L_0^2$) is less than 10%. It also should be noted that the relatively large error can be found near the first interface between two adjacent sections, owing to the drastic changes of width and the corresponding uneven pressure distribution [37].

It can also be seen that liquid front velocities in both the initial and narrower rectangular sections follow the Lucas-Washburn law while those in the control section do not. The velocity in the control section of all three elements is shown to vary linearly, including acceleration, deceleration, and uniform velocity, shown by red, blue, and green colors in Fig. 2(b), respectively. These constant velocity gradients and the corresponding closely matched simulation results indicate that the basic component is able to control the velocity in capillary flow processes. Therefore, it can be adopted to facilitate the design of functional capillary elements, as demonstrated in the following section.

## 6. Controllable capillary flow

Precise control of capillary flow in porous media is of growing interest for both scientists and engineers, with broad applications in the field of capillary-driven devices [7]. Previous works focus mainly on fluid flow in channels or tubes, but not much work has dealt with porous media. Here we consider a homogeneous porous medium with a variable cross-sectional profile to realize the control of the flow velocity. We demonstrate the design of two types of capillary elements with specific functions. By integrating the functional capillary elements, fluidic devices to achieve a simple example of controllable capillary flow are obtained.

### 6.1 Functional capillary elements

Inspired by the recent studies on capillary elements in channels or tubes [39-40], we demonstrate the design two types of functional capillary elements, i.e., flow amplifiers and flow resistors, based on the capillary flow in porous media, to achieve specific objectives of flow regulation.

### I) Capillary flow amplifier

For capillary-based microfluidic devices, such as chemical reactors and detectors, a controllable and constant flow velocity is usually needed, and within a given device different flow velocities may be required within a single porous medium to realize the function of grading reaction or detection. To this end, inspired by a typical electrical element, amplifier, to increase the magnitude of a signal, a fluid flow amplifier can be introduced. More specifically, flow amplifiers can be used to tailor the capillary flow, imparting different uniform velocities in individual sections of a multi-sectional porous sample. Here, in a generalized concept, capillary flow amplifiers are designed not only to increase but also to decrease flow velocities.

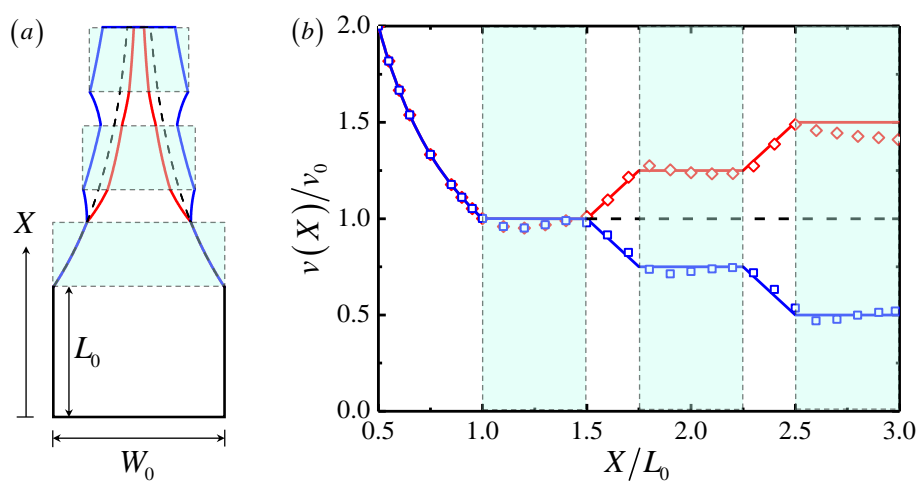

**Figure 3.** (a) Designed shape profiles of capillary flow amplifiers with the function of increasing (red) or decreasing (blue) the liquid front velocity; (b) Comparison of numerical simulations (symbols) and theoretical predictions (lines) of liquid front velocity versus position along flow direction for both two amplifiers.

Two amplifiers are designed to either increase or decrease flow velocity in a stepwise manner (i.e., velocity gradients at control sections in the amplifiers are set as $v_1 = v_0$, $v_3 = (1 \pm 1/4)v_0$ and $v_5 = (1 \pm 1/2)v_0$, where the subscript represents the position in series of the variable section), the geometric shape profiles are shown with red and blue colors respectively. For each amplifier, adjacent control sections with constant flow velocity are connected by an accelerating (or decelerating) section to maintain the continuous transition of velocity. Correspondingly, the velocity gradient at each section is calculated as $a_1 = a_3 = a_5 = 0$ and $a_2 = a_4 = \pm D/L_0^2$ for the cases of increasing or decreasing flow, respectively. As a reference, the profile of the sample with continuous constant velocity is also included by a black dotted line.

The theoretical predictions of the variation of liquid front velocity with position along the flow direction in the designed amplifiers are plotted in Fig. 3(b) as lines where the two cases with gradually increasing and decreasing velocities are marked with different colors. For comparison, the results of the corresponding FEA simulations for both amplifiers are also included and represented by symbols. It is clearly seen that the numerical simulations match closely with the theoretical predictions for both cases. Similar to the flow regulators designed here, a larger error is found at the junction of adjacent sections due to the sudden changes in width. These novel functional devices provide the possibility of piecewise controlling flow velocity in capillary-driven fluidic devices, with implications for practical applications, such as chemical reactors based on microfluidics, which require precise control of localized flow velocities.

**II) Capillary flow resistor**

Capillary flow resistors, i.e., elements used to reduce the capillary flow velocity, have been widely applied in capillary-driven microsystems [40]. Existing designs of flow resistors focused on capillary flow in tubes or channels [38, 40], while the idea of a flow resistor based on porous media has yet to be explored. Here we demonstrate a flow resistor device based on 2D porous media. The resistor considered here involves both positive (i.e., decreasing the flow rate) and negative (i.e., increasing the flow rate) resistances. In contrast to flow amplifiers, in these elements, the velocity is not uniform. Rather than tuning the velocity of the liquid front, control of total flow time is achieved by tailoring successive acceleration and deceleration stages.

Figure 4(a) shows the geometric profiles of the capillary flow resistors with positive and negative resistances. The resistors are designed using the methodology developed in Section 3, which is based on multi-sectional porous media. The control section consists of two adjacent accelerating and decelerating components with a constant velocity gradient and is sandwiched between two sections with constant velocity. For the designed resistors, from the red one to the blue one, the velocity gradients at the two sections in the control region are $a = \pm 3D/4L_0^2$, $\pm D/2L_0^2$, $\mp D/2L_0^2$ and $\mp 3D/4L_0^2$, respectively. For reference, the profile of a sample with constant velocity is included and shown as a black dotted line. Theoretical predictions and numerical simulations of liquid front velocities versus position along the flow direction for four resistors are plotted in Fig. 4(b). Good agreement is found between simulations and predictions for all four cases. Relative errors are found to be greater in positive resistors than for negative resistors. It also should be noted that the total flow duration in the control section can be tuned by changing the resistance, allowing the design of flow elements with controlled residence time.

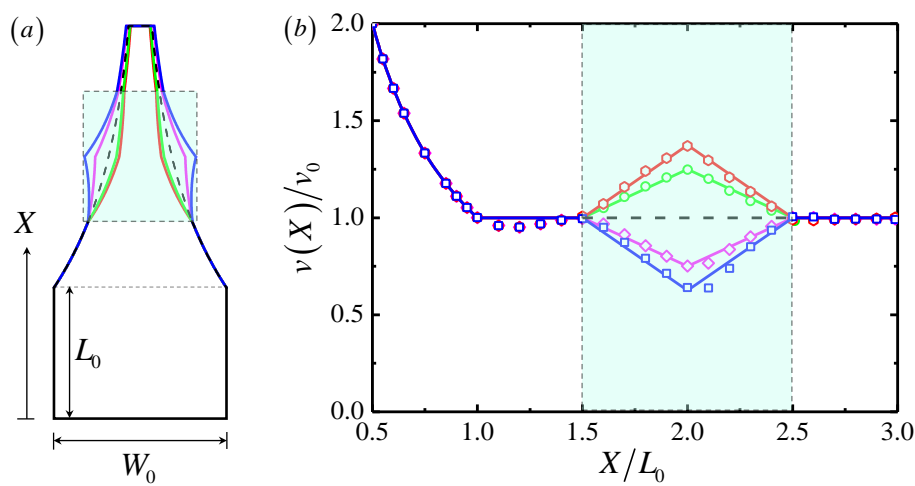

**Figure 4.** (a) Designed profiles of capillary flow resistors yielding positive and negative resistances; (b) Comparison of numerical simulations (symbols) and theoretical predictions (lines) of liquid front velocity versus position for all four resistors.

For the applications in microfluidic and diagnostic devices, the residence time of a fluid or analyte in a given part of the device is of critical importance [32]. For a flow velocity of $v_0$, by introducing a dimensionless parameter, $k$ which is also be called resistance coefficient, the peak velocity in the control section of a resistor can be expressed as $v_p = (k+1) \cdot v_0$. It is clear that the parameter $k$ is related to the velocity gradient in the control section and can be used to characterize the resistance of flow resistors. Specifically, when $k < 0$, the resistor presents positive resistance, while negative resistance will be found when $k > 0$. As a special case, $k = 0$ means zero resistance, and a constant velocity $v_0$ through the control section is obtained. Generally, the range of parameter $k$ can be from −1 to infinity. However, when the value of $k$ is much larger or smaller than 0, the geometric shape profile will be too sharp and difficult to achieve. Thus, here we only consider the range of $k$ between to −0.5 and 0.5.

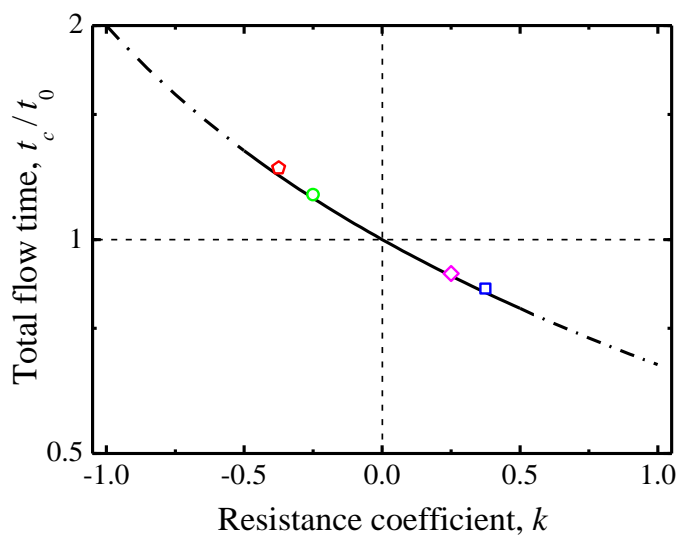

**Figure 5.** The normalized total flow time in the control section of capillary flow resistors $t_c/t_0$ as a function of the resistance coefficient $k$.

When the length of the control section is $l_c$, the total flow time in this section of the reference sample is calculated as $t_0 = l_c/v_0$. Accordingly, the total flow time in the control section of the resistors can be obtained as $t_c = 2/(2+k) \cdot t_0$. Following this correlation, the normalized total flow time, $t_c/t_0$, can be plotted as a function of the resistance coefficient $k$, as shown in Fig. 5. The simulated results of four designed resistors are also included for the purpose of comparison and are consistent with the theoretical results. It is clear that the resistance of the resistors can be tuned by changing the value of $k$, i.e., tuning the velocity gradient via changing the geometric shape of the control section in the resistors.

In this section, two novel capillary elements, i.e., flow amplifier and flow resistor, are shown. To facilitate the design of capillary flow devices, we introduce a symbolic representation of capillary elements with their corresponding features and functions (see Table 1). We will further demonstrate the design of a controllable flow device by integrating these functional capillary elements.

**Table 1**. Symbolic representations, flow features, and function of the designed capillary elements including (i) flow amplifier and (ii) flow resistor.

| | Symbolic | Flow feature | Function |
|---|---|---|---|
| Amplifier | 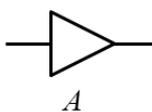 | 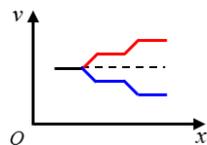 | Control of constant velocities at different sections |

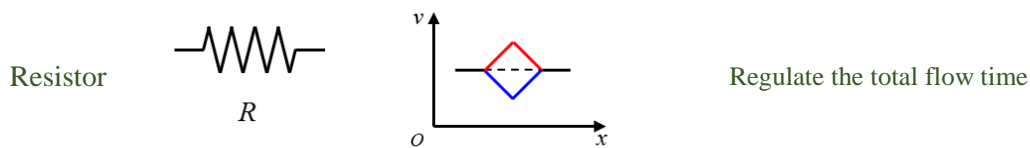

### 6.2 Controllable capillary flow in integrated device

Controllable (or programmable) flow in capillary force driven microfluidic networks has drawn attention towards applications in improved diagnostic devices [38, 45]. Most controllable fluidic devices are designed using multiple disjunct components, and only a single function can be achieved in a separate device [46]. We demonstrate here the design of a continuous flow device with multi-functions based on porous media to achieve controllable capillary flow.

As a simple demonstration of controllable capillary flow in porous media, a sample with integrated capillary elements is presented here. The geometric shape of the designed sample is shown in Fig. 6(a), and combines two functional elements i.e., a resistor ($a = -D/2L_0^2$ and $D/2L_0^2$) shown by blue lines, appended by an amplifier ($a = 3D/2L_0^2$ and 0) shown by green lines. A basic component is contained at the front of the resistor to supply a constant velocity ($a = 0$) as a reference, as shown by red lines. For each variable section, the geometric shape profile is calculated from the obtained theoretical model (13).

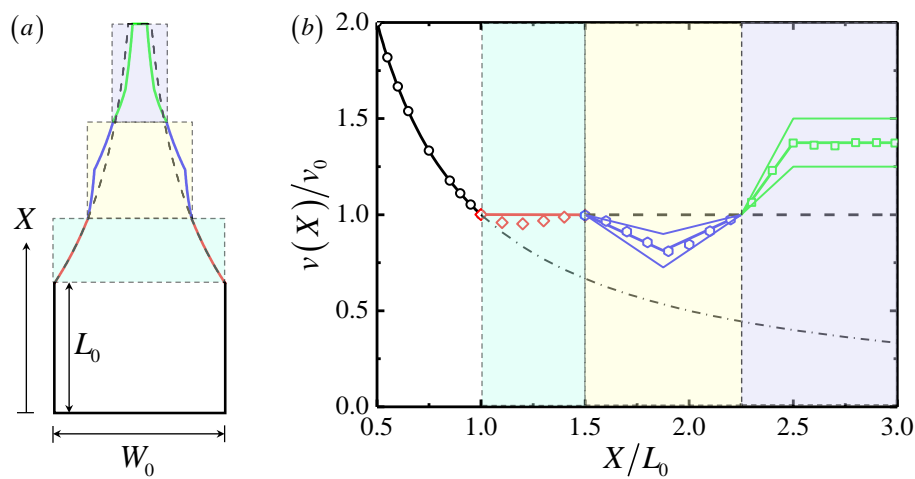

**Figure 6.** (a) Designed shape profiles of porous sample with programmable capillary flow; (b) Comparison of numerical simulations (symbols) and theoretical predictions (lines) of liquid front velocity versus position along flow direction of porous sample.

By employing the proposed numerical method, the capillary flow process in the sample is simulated, and the variation of liquid front velocity with position is plotted in Fig. 6(b) as symbols, with the theoretical results included as solid lines. The theoretical design of the controllable capillary flow device is verified by the good agreement between numerical and theoretical results. By integrating the designed functional capillary elements, we can realize

https://doi.org/10.1016/j.jcis.2018.12.068    16

controllable flow processes in porous media. It is clear that the designed flow process is deviated from the classical Lucas-Washburn law corresponding to specimens of uniform width, as shown by the dash-dotted line. The flow process can be further tuned by changing the parameters of each component. As shown in Fig. 6(b) with thin solid lines, the total flow time can be tuned by the resistor element, and the flow velocity at each stair can be tuned as required by the amplifier element. Additionally, more complex flow processes can be achieved by combining multiple elements.

## 7. Extension to 3D porous structures

In the previous analysis, we considered only the structure with sections of constant thickness, see Fig. 7(a), which can be treated as a 2D structure. For some applications, the cross-sectional shape needs to be changed in two dimensions, e.g., the circular or rectangular cross section changes along the flow direction, as shown in Fig. 7(b) and (c), and the porous sample should be treated as a 3D structure. Such capillary flow in three dimensions is naturally of great significance across diverse engineering applications [47, 48]. Even though the proposed theoretical model (i.e., Eq. (13)) is derived from a 2D geometry and based on the 1D liquid front assumption, it can nevertheless be extended to describe the flow process in a 3D structure [37].

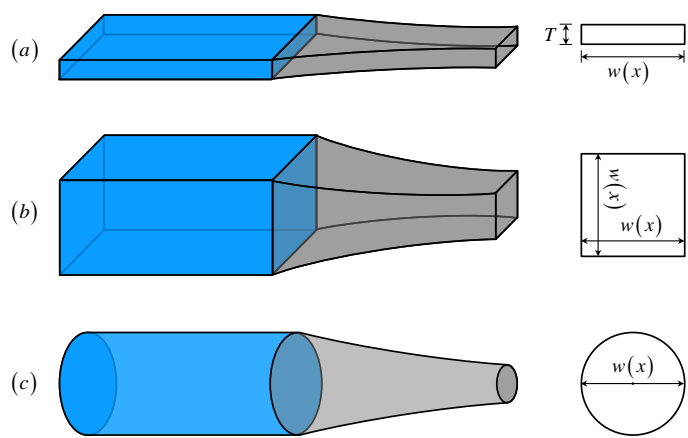

**Figure 7.** Sketch of designed programmable capillary flow samples based on porous media: (a) 2D structure; (b) 3D structure with rectangular cross section; and (c) 3D structure with circular cross section.

The theoretical model presented in Section 3 (Eq. (13)) describes the variation of the width of a 2D porous sample with a constant thickness. For a 3D porous structure, if the cross-sectional shape changes in two dimensions proportionately, the variation in each dimension corresponds to the square root of the variation of the cross-sectional area. That means, the general variation of cross-sectional size for both 2D and 3D porous structures, i.e., the width or thickness for the 2D case in Fig. 7(a), and the radius or edge length for 3D cases in Figs. 7(b) and (c), can be expressed as

$$\frac{w_i(x_i)}{W_{i-1}} = \mathrm{Exp}\left[-\frac{1}{n}\left(\frac{x_i}{L_{B_{i-1}}} + \frac{1}{2} \cdot \frac{1-\alpha_i}{\alpha_i} \cdot \frac{x_i^2}{L_{B_{i-1}}^2}\right)\right] \Bigg/ \left(1 + \frac{1-\alpha_i}{\alpha_i} \cdot \frac{x_i}{L_{B_{i-1}}}\right)^{\frac{1}{n}} \quad (19)$$

where $n$ indicates the number of the variable dimensions. $n = 2$ means that the cross section of the porous sample is varied in two dimensions, and $n = 1$ means only one dimension of the cross section will change. In addition, when $n = 1$, Eq. (19) will reduce to Eq. (13), with respect to the 2D special case. By employing this general equation, controllable capillary flow in 3D porous structures can be designed systematically.

## 8. Discussions and Conclusions

In this work, a novel approach is introduced whereby the geometry of porous structures is tailored to achieve controllable capillary flow. Through a methodology based on Darcy's law, we designed a basic capillary flow component to maintain a constant flow velocity gradient, in which the flow processes deviate from the Lucas-Washburn law [10, 11]. This component design provides greater versatility relative to earlier approaches which allowed only for constant velocity or flux ( flow rate) in porous media [42, 43]. The elements designed here offer a new pathway to either increase or decrease flow velocity in a well-controlled manner, enabling the achievement of complex flow distributions.

Based on the proposed basic component, two types of novel capillary flow elements, i.e., flow amplifiers and flow resistors, are developed. Different from the previous designs consisting of channels or tubes [39-40], the current capillary elements are based on porous media, and only the geometrical shape is varied to control flow. The flow features of these capillary elements and their advantages with respect to the regulation of flow processes have been systematically discussed. It is found that both the local flow velocity and the total flow time can be precisely regulated in specific control segments of the porous structures, with implications towards diagnostic and microfluidic devices.

We further demonstrate the design of a porous media based multi-functional device that integrates the designed capillary elements to achieve controllable capillary flow with broad adjustability, which is made as a single unit rather than a flow circuit consisting of many separate parts [38, 46]. To expand the application range, the extension of the theoretical methodology to 3D cases is also discussed. In addition, the realizability of these theoretical designs are verified by simulations based on the developed numerical method. In the future studies, the capillary elements developed here can be combined in greater numbers to yield porous devices with yet more complex flow features. The precise tuning of the flow process can be further enhanced by considering external effects such as evaporation [19, 20], gravity [21], and swelling of the porous matrix [22, 23].

Liu, M., Suo, S., Wu, J., Gan, Y., Hanaor, D. A., & Chen, C. Q. (2019). Tailoring porous media for controllable capillary flow. *Journal of colloid and interface science*, *539*, 379-387.

## Acknowledgments

The authors are grateful for the financial support from the Science Challenge Project of China (Grant No. TZ2018007), and the National Natural Science Foundation of China (No. 11472149 and 11732007). M.L. acknowledges support from the Endeavour Research Fellowship founded by Australian Government. Y.G. acknowledges support from The University of Sydney SOAR Fellowship.

**Competing financial interests:** The authors declare no competing financial interests.